\documentclass[aps,prl,reprint,groupedaddress,floatfix]{revtex4-1}
\usepackage{graphicx}
\usepackage{epsfig}

\begin{document}


\title{The percolation critical polynomial as a graph invariant}


\author{Christian R. Scullard}
\email[]{scullard1@llnl.gov}
\affiliation{Lawrence Livermore National Laboratory, Livermore CA 94550, USA}


\date{January 8, 2013}

\begin{abstract}
Every lattice for which the bond percolation critical probability can be found exactly possesses a critical polynomial, with the root in $[0,1]$ providing the threshold. Recent work has demonstrated that this polynomial may be generalized through a definition that can be applied on any periodic lattice. The polynomial depends on the lattice and on its decomposition into identical finite subgraphs, but once these are specified, the polynomial is essentially unique. On lattices for which the exact percolation threshold is unknown, the polynomials provide approximations for the critical probability with the estimates appearing to converge to the exact answer with increasing subgraph size. In this paper, I show how this generalized critical polynomial can be viewed as a graph invariant, similar to the Tutte polynomial. In particular, the critical polynomial is computed on a finite graph and may be found using the recursive deletion-contraction algorithm. This allows calculation on a computer, and I present such results for the kagome lattice using subgraphs of up to $36$ bonds. For one of these, I find the prediction $p_c=0.52440572...$, which differs from the numerical value, $p_c=0.52440503(5)$, by only $6.9 \times 10^{-7}$.
\end{abstract}


\maketitle

\section{Introduction}
\begin{figure}
\includegraphics{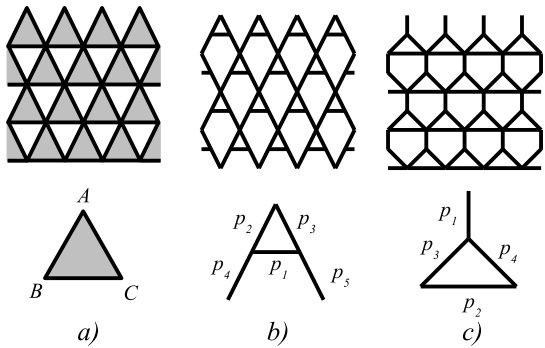}
\caption{a) A 3-uniform hypergraph. The shaded region need not be a simple triangle; b) the martini-A lattice; c) the martini-B lattice.}
\label{fig:3uniform}
\end{figure}

Percolation is the study of the formation of random clusters on lattices. Given an infinite lattice, we declare each bond to be open with probability $p$, and closed with probability $1-p$. The resulting random clusters grow in average size with $p$ until we reach the critical point, $p_c$, above which an infinite cluster appears. The determination of $p_c$ is an unsolved problem except in one dimension and on two-dimensional lattices that are self-dual 3-uniform hypergraphs \cite{BollobasRiordan11}. An example is shown in Figure \ref{fig:3uniform}a, where the shaded triangle, excluding corner vertices, can represent any configuration of bonds, sites and correlations. Critical thresholds on these lattices are given by the Ziff criterion \cite{Ziff06},
\begin{equation}
 P(\bar{A},\bar{B},\bar{C})-P(A,B,C)=0 \label{eq:ziff}
\end{equation}
where $P(A,B,C)$ is the probability that all three vertices are connected through open paths within the triangle, and $P(\bar{A},\bar{B},\bar{C})$ is the probability that none are connected. Application of (\ref{eq:ziff}) to find a bond threshold results in a polynomial in $p$, with order at most equal to the number of bonds in the unit triangle. We may consider each bond of an $n-$bond triangle to have a different probability, giving a critical surface of the generic form
\begin{equation}
 f(p_1,...,p_n) = 0
\end{equation}
where $f$ is at most first order in any of its arguments, a property referred to as multi-linearity. Examples of such critical surfaces include the square lattice,
\begin{equation}
 S(p_1,p_2)=1-p_1-p_2, \label{eq:square}
\end{equation}
the honeycomb lattice,
\begin{equation}
 H(p_1,p_2,p_3)=1 - p_1 p_2 - p_1 p_3 - p_2 p_3 + p_1 p_2 p_3. \label{eq:hex}
\end{equation}
and the triangular lattice,
\begin{equation}
 T(p_1,p_2,p_3)=1 - p_1 - p_2 - p_3 +p_1 p_2 p_3\ . \label{eq:tri}
\end{equation}
The critical probability for the lattice is then found by setting all probabilities equal. For the square lattice, for example, this gives the polynomial $S(p,p)=1-2p=0$, and the threshold $p_c=1/2$.

In this paper, I will show how such critical surfaces may be generalized to any lattice, even those which are not in the solvable class, by employing a deletion-contraction algorithm. Such surfaces may not be exact, but we will see that they give approximations that, in principle, can be made arbitrarily precise. I begin by describing the deletion-contraction property of critical surfaces derived from (\ref{eq:ziff}). Then I show how this can be used to define the critical surfaces and polynomials for unsolved problems. Although it appears that the definition may not result in a unique critical surface in all cases, I present an argument to show that it is in fact always well-defined. I conclude by reporting generalized critical polynomials for the kagome lattice, giving approximations to the threshold that become increasingly more precise, with the best estimate within only $6.9 \times 10^{-7}$ of the numerically determined value. Although I use the kagome lattice as an illustrative example, this procedure can be applied on any periodic lattice.

\section{Deletion-contraction}
Consider the martini-A lattice, with the probability assignments shown in Figure \ref{fig:3uniform}b. If we set $p_1=1$, this bond is contracted and its end vertices merged, the result being the honeycomb lattice with two bonds doubled in parallel. This doubled bond can be replaced by a single effective bond, so we have
\begin{equation}
 H(1-[1-p_2][1-p_3],p_4,p_5) .
\end{equation}
By setting $p_1=0$, we delete the bond, and the result is the square lattice with bonds doubled in series. Thus, we have,
\begin{equation}
 S(p_2 p_4,p_3 p_5) .
\end{equation}
The critical surface of the martini-A lattice is given by (\ref{eq:ziff}). The only way to reduce to the correct deleted and contracted surfaces and preserve the required multi-linearity property is to set
\begin{eqnarray}
A(p_1,p_2,&p_3&,p_4,p_5)= \cr
&p_1& H(1-[1-p_2][1-p_3],p_4,p_5) \cr
&+&(1-p_1) S(p_2 p_4,p_3 p_5), \label{eq:A}
\end{eqnarray}
which has the formal appearance of an average of the two special cases. It is a basic property of critical surfaces found using (\ref{eq:ziff}) that they satisfy such deletion-contraction formulas. That is, if such a lattice, $L$, has a triangular unit of $n$ bonds such that deletion and contraction of the $p_1$ bond results in the lattices $L_0$ and $L_1$, then we have for the critical surface of $L$,
\begin{eqnarray}
 L(p_1,&p_2&,...,p_n) = p_1 L_1(p_2,p_3,...,p_n)\cr
&+&(1-p_1) L_0(p_2,p_3,...,p_n) . \label{eq:L}
\end{eqnarray}
As another example, we find for the martini-B lattice (Figure \ref{fig:3uniform}c),
\begin{eqnarray}
 B(p_1,p_2,p_3,p_4)&=&p_4 S(p_1,1-[1-p_2][1-p_3]) \cr
&+& (1-p_4) S(p_1 p_3,p_2) . \label{eq:B}
\end{eqnarray}

\section{Generalized critical surfaces and polynomials}
The deletion-contraction formula can be used to extend the definition of the critical surface to lattices for which the threshold is not known exactly, and in these cases it will be referred to as the {\it generalized} critical surface. For the kagome lattice of Figure \ref{fig:kagome2}a, an unsolved problem, we may write,
\begin{eqnarray}
 K(p_1,&p_2&,p_3,p_4,p_5,p_6)= \cr
&p_4& B(1-[1-p_5][1-p_6],p_1,p_2,p_3) \cr
&+&(1-p_4) A(p_1,p_2,p_3,p_5,p_6), \label{eq:fullK}
\end{eqnarray}
where $A$ and $B$ are given by (\ref{eq:A}) and (\ref{eq:B}). Generalized critical polynomials are found by setting all probabilities equal, and in this case $K(p,p,p,p,p,p)=0$, i.e.,
\begin{equation}
 1-3 p^2 - 6 p^3 + 12 p^4-6 p^5 + p^6 =0 \label{eq:K}
\end{equation}
which gives Wu's well-known \cite{Wu79,ScullardZiff06} approximation $p_c=0.52442971...$, compared to the numerical \cite{Feng08} $p_c=0.52440502(5)$. Thus, a generalized critical polynomial may be defined on any lattice for which known lattices appear upon deletion and contraction of a bond, but there is no guarantee that the resulting prediction will be exact. In fact, the definition extends to any periodic lattice. If $L$ is in the solvable class, the polynomial is given by (\ref{eq:ziff}), and otherwise it is defined by (\ref{eq:L}), with this formula applied recursively if either $L_0$ or $L_1$ have unknown threshold. By repeated application, known lattices eventually appear and the full critical surface can be found. That it satisfies such a deletion-contraction formula makes the generalized critical surface similar to other graph invariants, such as the Tutte polynomial and its special cases (e.g., the chromatic polynomial) \cite{BollobasBook}.

In the following, I call the subgraph of a lattice, on which different probabilities are assigned, the ``base'' of the process, which is then tiled to form the infinite lattice. For example, in Figures \ref{fig:kagome2}a, \ref{fig:kagome2}b, \ref{fig:kagome4} and \ref{fig:kagome6} are bases for the kagome lattice consisting of one, two, four, and six unit cells. I now enumerate some properties of generalized critical surfaces and polynomials:

\begin{enumerate}
\item They are unique once the base and tiling are specified \cite{Scullard11,Wu10}. In particular, critical surfaces are independent of the bonds chosen to delete and contract at each step. To prove this, we procede by induction. First, we call critical surfaces for which the final result is independent of deletion-contraction bond order {\it well-defined}. Clearly, any self-dual 3-uniform hypergraph has this property, as the polynomial is defined by (\ref{eq:ziff}); deletion-contraction must give this answer, regardless of the bond chosen, because it only imposes the correct boundary values on the critical surface. Similarly, any critical surface that is a special case of a well-defined surface is itself well-defined. Next, we show that the critical surface for a lattice, $L$, that is not a realization of a self-dual 3-uniform hypergraph, and is thus not solved by (\ref{eq:ziff}), is well-defined if every critical surface resulting from deleting and contracting any bond of $L$ is well-defined. The kagome lattice with the $6$-bond base is one example of such a system. Assuming that $L$ has this property, consider the generic formula (\ref{eq:L}). The question is whether this is the same as what we would obtain by choosing the $p_2$ bond instead,
\begin{eqnarray}
  \tilde{L}(p_1,&p_2&,...,p_n) =p_2 \tilde{L}_1(p_1,p_3,...,p_n)\cr
&+&(1-p_2) \tilde{L}_0(p_1,p_3,...,p_n) , \label{eq:Ltilde}
\end{eqnarray}
i.e., whether $L=\tilde{L}$. We can apply deletion-contraction on $p_2$ to the graphs $L_0, L_1$, and on $p_1$ to $\tilde{L}_0, \tilde{L}_1$ to obtain $L_{00}, L_{01}$, etc. We then have,
\begin{eqnarray}
 L&=& p_1 p_2 L_{11}(p_3,...,p_n) \cr
&+&p_1 (1-p_2) L_{10}(p_3,...,p_n) \cr
&+&(1-p_1) p_2 L_{01}(p_3,...,p_n) \cr
&+&(1-p_1)(1-p_2) L_{00}(p_3,...,p_n)
\end{eqnarray}
and
\begin{eqnarray}
 \tilde{L}&=& p_1 p_2 \tilde{L}_{11}(p_3,...,p_n) \cr
&+&p_1 (1-p_2) \tilde{L}_{10}(p_3,...,p_n) \cr
&+&(1-p_1) p_2 \tilde{L}_{01}(p_3,...,p_n) \cr
&+&(1-p_1)(1-p_2) \tilde{L}_{00}(p_3,...,p_n),
\end{eqnarray}
and clearly these two expressions are the same if $L_{11}=\tilde{L}_{11}$, $L_{10}=\tilde{L}_{10}$, etc. By assumption, $L_1$, $L_0$, $\tilde{L}_1$ and $\tilde{L}_0$ are well-defined. Now, the function $\tilde{L}_{11}$ arises by contracting the bonds $p_2$ and $p_1$ in $L$, which is the same way we arrive at $L_{11}$. Thus, these two functions must describe the same lattice with a well-defined critical surface, and so we have $\tilde{L}_{11}=L_{11}$. By the same reasoning, $L_{10}=\tilde{L}_{10}$, etc., and we conclude that if every critical surface arising from deleting and contracting any bond of $L$ is well-defined, then so is $L$. Because each step of the algorithm lowers the number of bonds in the base by (at least) one, well-defined surfaces eventually appear and the general result follows by induction. A slight complication arises due to the existence of certain seemingly pathological cases. Consider, for example, setting $p_1=0$ on the square lattice. Now we have created a one-dimensional system, with the trivial critical point $p_2=1$, or $1-p_2=0$. This latter form is the correct surface to use whenever the one-dimensional case appears, and is indeed reflected in the formula (\ref{eq:square}). A less straightforward example appears on the square lattice when we set $p_1=1$ in (\ref{eq:square}). It is not obvious how to interpret the result of contracting this bond since it seems to collapse the lattice. However, in such cases it is better to think of the bond as not contracting to zero length but simply carrying probability $1$ and thus introducing infinitely long linear clusters on the lattice. In the square case, the system is supercritical unless we also set $p_2=0$, which gives a critical one-dimensional system. The supercritical phase of (\ref{eq:square}) is represented by $S(p_1,p_2)<0$, so by the above considerations we would expect the surface for $p_1=1$ to read $-p_2=0$, which is indeed what we find from (\ref{eq:square}). Even if the lattice is unknown, by keeping these ``supporting'' bonds uncontracted when they are used in the algorithm, we can continue until we reach cases on which we may use (\ref{eq:ziff}), and then the supporting bond is just given probability $1$ in the formula. Finally, consider the result of setting $p_1=p_2=1$ in the triangular formula (\ref{eq:tri}), thus imposing an infinite two-dimensional cluster. This system is super-critical (there is no infinite cluster at the critical point) regardless of the value of $p_3$, and thus we have $T(1,1,p_3)=-1$, which is a general rule for whenever such a situation arises. This completes the discussion of potentially problematic cases, and thus the proof that the critical surface for a general lattice and base is well-defined. When calculating the polynomials reported below, each one was computed several times using different bond paths to ensure that the code was working properly. For the largest bases of $36$ bonds, each calculation was repeated over a dozen times and the same polynomial always resulted.

\item If the single-cell prediction is not exact, it is generally very close, usually within $10^{-5}$ of the numerically determined threshold \cite{Scullard10}. It is not clear why this should be so, but it has been clearly demonstrated for many different systems, including all the Archimedean lattices \cite{Scullard10,Scullard11}. In addition, proper choices of larger bases lead to predictions closer to the exact answer, with accuracy increasing with base size. It is this latter conjecture I seek to support here by computing polynomials on the kagome lattice for bases of up to $36$ bonds.
\item Conversely, if the critical polynomial gives the exact threshold for a base of a single unit cell, as is the case for self-dual $3-$uniform hypergraphs, then any critical polynomial using a larger base makes the same prediction, i.e. the original polynomial always factors out. This is also a conjectured property but it would seem to be necessary for consistency, and I have found no counter-examples. While it would be ideal to have a lattice for which the exact threshold is known and is not a root of a polynomial, so that we can directly check the convergence of these polynomials, no such example is known at present and the conjecture can only be tested against numerically determined values.
\item In many cases, if the single-cell polynomial does not give the exact answer, then it can be shown \cite{Scullard11} that no critical polynomial derived from a finite-sized base will give the correct percolation threshold. This is true for the kagome lattice, as discussed below.
\end{enumerate}

Generalized critical polynomials have been found for all the Archimedean lattices \cite{Scullard08,Scullard10,Scullard11}. The polynomial (\ref{eq:K}) was originally found by Wu using his ``homogeneity'' \cite{Wu79} assumption and recently, he extended the method to include kagome subnets \cite{Wu10}, with excellent results \cite{Ding10}. In fact, the homogeneity approximation gives a prediction for the full $q-$state Potts critical frontier, but at present it appears limited to the kagome lattice and its subnets as it relies on a transformation from the triangular lattice to these kagome-type graphs. Although I focus on the kagome lattice in this paper, mostly because of the interest it has attracted over the years (e.g., \cite{ZiffGu,Tsallis,Wu79}), the method presented here may be applied on any periodic lattice for either site or bond percolation.

\section{Kagome lattice polynomials}
In previous work, critical polynomials were found ``by hand'', with the $(4,6,12)$ lattice and its $18-$bond unit cell probably representing the limit of what one would be inclined to do this way (see the appendix of \cite{Scullard11}). However, the recursive nature of the algorithm makes it an ideal problem for a computer, and here I report the results of using a program to calculate critical polynomials on the kagome lattice for bases containing $12$, $24$, and $36$ bonds. At each step, the program chooses a bond, and finds the two graphs that result from its deletion and contraction. It knows a small number of graphs (e.g., triangular, honeycomb, and square lattices), and it repeats the deletion-contraction algorithm recursively until it recognizes all the lattices it has found. The output is a set of function definitions, like equations (\ref{eq:A}), (\ref{eq:B}) and (\ref{eq:fullK}) which can be evaluated in a computer algebra package to get the critical surface and then the critical polynomial. Of course, there are many issues to overcome in programming this scheme, and a full account is given elsewhere \cite{Scullard12}.

\subsubsection{Base of 2 unit cells}
\begin{figure}
\includegraphics{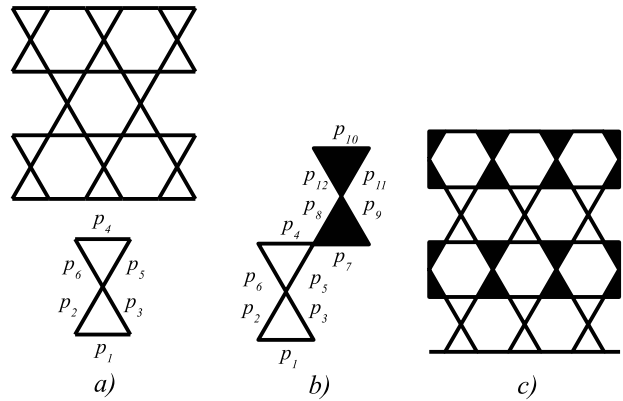}
\caption{a) The kagome lattice and its single unit cell $6$-bond base; b) a $12$-bond base for the kagome lattice; c) a tiling for the base in b).}
\label{fig:kagome2}
\end{figure}

The first extension beyond a single unit cell base is that shown in Figure \ref{fig:kagome2}b, in which we employ a base using two unit cells which are indicated by different colors, with the tiling shown in Figure \ref{fig:kagome2}c. The polynomial can be written in the factored form
\begin{eqnarray}
&-&(1 - 3 p^2 - 6 p^3 + 12 p^4 - 6 p^5 + p^6) \nonumber \\
&\times& (-1 - p^2 - 2 p^3 + 10 p^4 - 10 p^5 + 3 p^6) .
\end{eqnarray}
We recognize the first term in brackets as the polynomial for the $6$-bond base. The second term has no real roots, and thus the prediction here is the same as the one we found previously, $p_c=0.52442971...$ . There is no other way to tile this base to give a different result.

\subsubsection{Bases of 4 unit cells}
\begin{figure}
\includegraphics{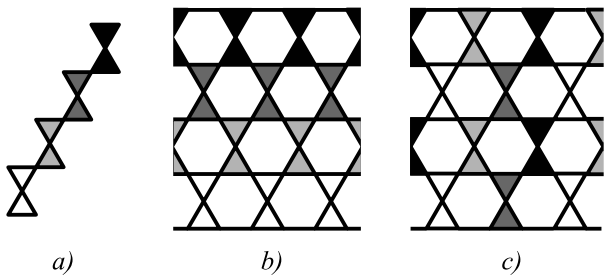}
\caption{A $24$-bond base for the kagome lattice with two inequivalent embeddings.}
\label{fig:kagome4}
\end{figure}

Using the base consisting of $4$ unit cells shown in Figure \ref{fig:kagome4}a, we may tile it in two different ways as indicated in Figures \ref{fig:kagome4}b and \ref{fig:kagome4}c. Starting with Figure \ref{fig:kagome4}b, the critical polynomial can be written in the factored form
\begin{eqnarray}
&-&(1 - 3 p^2 - 6 p^3 + 12 p^4 - 6 p^5 + p^6) \times \cr
& &(-1 - p^2 - 2 p^3 + 10 p^4 - 10 p^5 + 3 p^6) \times \cr
& &(1 - 2 p^2 - 4 p^3 + 7 p^4 + 24 p^5 - 28 p^6 - 64 p^7 + \cr
& &   172 p^8 - 184 p^9 + 110 p^{10} - 36 p^{11} + 5 p^{12})
\end{eqnarray}
Once again, the first term in brackets is just the $6$-bond polynomial (\ref{eq:K}), and the others have no roots in $[0,1]$. The prediction is again the same as the $6$-bond estimate.

Things finally become more interesting when we use the staggered embedding in Figure \ref{fig:kagome4}c. In this case, the polynomial is
\begin{eqnarray}
1 &-& 6 p^4 - 24 p^5 - 24 p^6 - 24 p^7 + 27 p^8 \nonumber \\ 
&+& 552 p^9 + 1056 p^{10} - 1224 p^{11} - 8548 p^{12} \nonumber \\ 
&-& 4872 p^{13} + 68568 p^{14} - 50664 p^{15} \nonumber \\
&-& 226650 p^{16} + 643944 p^{17} - 843684 p^{18} \nonumber \\
&+& 684384 p^{19} - 368886 p^{20} + 133152 p^{21} \nonumber \\ 
&-& 31068 p^{22} + 4248 p^{23} - 259 p^{24} = 0,
\end{eqnarray}
and the solution on $[0,1]$ is $p_c=0.52440672...$, which differs from the numerical result by $1.7 \times 10^{-6}$, a great improvement over the single-cell $6$-bond case. Although finding all the bases that give different polynomials is something of an art, especially as the number of bonds becomes large, there does not appear to be one that gives a different prediction using $24$ bonds.
\subsubsection{Bases of 6 unit cells}
There are many options for bases of $36$ bonds, and I will discuss only three here. If we take the 6-unit cell version of the base shown in Figure \ref{fig:kagome4}a with the embedding analogous to Figure \ref{fig:kagome4}b, we once again find a polynomial in which (\ref{eq:K}) appears as a factor. Thus, for this base and embedding, we get the $6$-bond estimate again. It is tempting to assume that this trend continues for larger bases and embeddings of this type.

\begin{figure}
\includegraphics{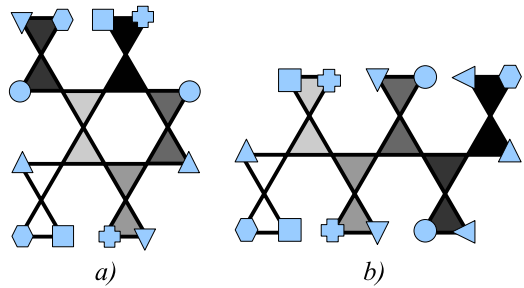}
\caption{$36$-bond bases for the kagome lattice. Matching shapes on the external vertices indicate how each base is tiled to create the lattice.}
\label{fig:kagome6}
\end{figure}

Turning now to the base shown in Figure \ref{fig:kagome6}a, in which the embedding is indicated by the matching shapes on the external vertices, we get the polynomial,
\begin{eqnarray}
1 &-& 3 p^4 - 12 p^5 - 20 p^6 - 60 p^7 - 132 p^8 + 56 p^9 \cr
&+& 684 p^{10} + 1440 p^{11} + 2108 p^{12} + 2052 p^{13} \cr
&-& 10452 p^{14} - 68708 p^{15} - 82980 p^{16} + 280152 p^{17} \cr
&+& 1316026 p^{18} - 49980 p^{19} - 12878976 p^{20} \cr
&+& 5124684 p^{21} + 90816816 p^{22} - 199458252 p^{23} \cr 
&-& 12979085 p^{24} + 816398808 p^{25} - 1939348056 p^{26} \cr
&+& 2677229528 p^{27} - 2575935942 p^{28} \cr
&+& 1832168220 p^{29} - 984362272 p^{30} \cr
&+& 400507236 p^{31} - 121897767 p^{32} + 26954680 p^{33} \cr
&-& 4096134 p^{34} + 382956 p^{35} - 16617 p^{36}=0, \label{eq:K6_1}
\end{eqnarray}
which has solution on $[0,1]$ $p_c=0.52440607...$, differing from the numerical value by $1.1 \times 10^{-6}$. A different base and embedding is shown in Figure \ref{fig:kagome6}b and has polynomial,
\begin{eqnarray}
1 &-& 6 p^4 - 24 p^5 - 14 p^6 + 36 p^7 + 39 p^8 - 100 p^9 \cr
&-& 462 p^{10} + 780 p^{11}+ 4583 p^{12} + 4812 p^{13} \cr
&-& 9276 p^{14} - 71600 p^{15} - 85626 p^{16} \cr
&+& 312336 p^{17} + 1091146 p^{18} - 509340 p^{19} \cr
&-& 9675936 p^{20} + 5297340 p^{21} + 66607704 p^{22} \cr
&-& 151097304 p^{23} - 5319734 p^{24} + 610494828 p^{25} \cr
&-& 1461237180 p^{26} + 2022998000 p^{27} \cr
&-& 1949295060 p^{28} + 1387593528 p^{29} \cr
&-& 745850356 p^{30}+ 303533928 p^{31} \cr
&-& 92388675 p^{32} + 20427736 p^{33} - 3103578 p^{34} \cr
&+& 290052 p^{35} - 12579 p^{36}=0,
\end{eqnarray}
with solution in $[0,1]$, $p_c=0.52440572...$, slightly better than the previous estimate and within $6.9 \times 10^{-7}$ of the numerical value.

\section{Discussion}
\begin{figure}
\begin{center}
\includegraphics{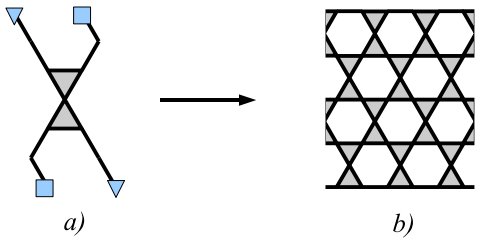}
\caption{a) The subgraph and embedding resulting from deleting $22$ bonds in Figure \ref{fig:kagome6}a, giving a single cell of the kagome lattice plus some connecting bonds; b) contracting the connecting bonds gives the kagome lattice partitioned into unit cells.}
\label{fig:kagomereduce}
\end{center}
\end{figure}

Clearly, we are observing the predictions converging to the exact answer with increasing base size. The results suggest that, if a generic base can be described as containing $n \times m$ unit cells, the exact answer is approached only as both $n$ and $m$ go to $\infty$ since the $1 \times n$ bases appear all to provide the same incorrect prediction. From these few examples it also appears that the kagome estimates are converging from above, as they all seem to be greater than the numerical value. However, I know of no argument that guarantees this trend will continue. Also, as the polynomials respect duality, i.e., making the substitution $p \rightarrow 1-p$ gives a polynomial for the dual graph, the estimates for the dual of the kagome lattice, the dice lattice, would converge from below. Note also that no polynomial for the kagome lattice derived from any finite-size base will give the exact answer. This can be seen as follows. Using the base in Figure \ref{fig:kagome6}a, we delete many of the bonds to leave a single unit cell and a few connecting bonds, as in Figure \ref{fig:kagomereduce}a. Now, we may contract the connecting bonds to recover the kagome system in which the base is a single cell, as in Figure \ref{fig:kagomereduce}b. By uniqueness, the prediction for this case must be equation (\ref{eq:K}), which contradicts the prediction found by setting all bonds equal in the full critical surface, i.e. equation (\ref{eq:K6_1}). Thus, if the single-cell polynomial does not provide the exact threshold, and for the kagome lattice it does not, then, although we can get arbitrarily close by using ever larger bases, no finite critical polynomial derived in this way will ever solve the problem.

There are many possible directions for further study. Aside from more firmly establishing the various conjectures, it would be interesting to try to quantify the manner of convergence to the exact solution, perhaps through a version of finite-size scaling. Another avenue is the extension to three and higher dimensions. While the polynomials are well-defined by the deletion-contraction algorithm in any dimension, preliminary results seem to indicate that they are not as successful at predicting higher-D critical points. However, this will be the subject of future work.

I have presented critical polynomials for the kagome lattice, up to bases of $36$ bonds. This is the limit of the current implementation of the algorithm, as it becomes progressively more difficult to add bonds due to the exponential complexity. It is not uncommon for a large calculation to produce over a million function definitions. Nevertheless, there is room for improvement in the efficiency, as the rule used to choose the bond for deletion and contraction at each step can have a significant effect on the rate of reduction to the known simple cases. I have hardly explored this issue and presently use what amounts to a random bond selection, rejecting only those choices that lead to undue complications (such as the contraction of supporting bonds described above). Moreover, this algorithm is perfect for a parallel implementation as it would require little inter-processor communication. It remains to be determined how much extra performance can be wrought from these considerations, but the problem will hopefully be seen as an interesting computational challenge.

\begin{acknowledgments}
I am grateful to Oliver Riordan for many valuable suggestions. I also thank Robert Ziff for the fruitful collaboration that led to this work, and Jesper Jacobsen for informative discussions. Finally, I thank an anonymous referee for providing several helpful comments. This work was performed under the auspices of the US Department of Energy by Lawrence
Livermore National Laboratory under Contract No. DE-AC52-07NA27344.
\end{acknowledgments}

\bibliography{scullard.bib}

\end{document}